\documentclass[amsmath,amssymb,floatfix,preprintnumbers,nofootinbib,prd,twocolumn,superscriptaddress]{revtex4}

\usepackage[utf8]{inputenc}
\usepackage{graphicx}

\usepackage[format=plain,justification=raggedright, singlelinecheck=false, font=small]{caption}
\usepackage[format=plain,justification=raggedright, singlelinecheck=false,font=small]{subcaption}

\usepackage{amssymb}
\usepackage{amsxtra}
\usepackage{amsmath}
\usepackage{booktabs,multirow,tabularx}
\usepackage{slashed}
\usepackage{float}
\usepackage{placeins}
\usepackage{rotating}
\usepackage{lscape}
\usepackage{color}
\usepackage{hyperref}
\usepackage{cancel}
\usepackage[Q=yes,pverb-linebreak=no]{examplep}
\usepackage{ulem}

\DeclareGraphicsExtensions{.pdf}
\graphicspath{{./figures/}}

\newcommand{\lsim}{\lesssim}

\newcommand{\eps}{\varepsilon}

\newcommand{\ord}[1]{\mathcal{O}{(#1)}}

\def\beq{\begin{equation}}
\def\bea{\begin{eqnarray}}
\def\eeq{\end{equation}}
\def\eea{\end{eqnarray}}
\def\beqal{\begin{align}}
\def\endal{\end{align}}

\definecolor{applegreen}{rgb}{0.55, 0.71, 0.0}
\definecolor{purple}{rgb}{0.5,0.0,0.5}

\makeatletter
\newcommand\footnoteref[1]{\protected@xdef\@thefnmark{\ref{#1}}\@footnotemark}
\makeatother

\DeclareFontFamily{U}{cbgreek}{}
\DeclareFontShape{U}{cbgreek}{m}{n}{
        <-6>    grmn0500
        <6-7>   grmn0600
        <7-8>   grmn0700
        <8-9>   grmn0800
        <9-10>  grmn0900
        <10-12> grmn1000
        <12-17> grmn1200
        <17->   grmn1728
      }{}
\DeclareFontShape{U}{cbgreek}{bx}{n}{
        <-6>    grxn0500
        <6-7>   grxn0600
        <7-8>   grxn0700
        <8-9>   grxn0800
        <9-10>  grxn0900
        <10-12> grxn1000
        <12-17> grxn1200
        <17->   grxn1728
      }{}

\makeatletter
\newcommand{\normalorbold}{%
  \ifnum\pdf@strcmp{\math@version}{bold}=\z@ bx\else m\fi
}
\makeatother

\begin{document}

\title{Displaced Signals of Hidden Vectors at the Electron-Ion Collider}

\author{Hooman Davoudiasl\footnote{email: hooman@bnl.gov}}

\affiliation{High Energy Theory Group, Physics Department,Brookhaven National Laboratory, Upton, NY 11973, USA}

\author{Roman Marcarelli\footnote{email: roman.marcarelli@colorado.edu}}

\author{Ethan T. Neil\footnote{email: ethan.neil@colorado.edu}}

\affiliation{Department of Physics, University of Colorado, Boulder, Colorado 80309, USA}

\begin{abstract}

The Electron-Ion Collider (EIC) provides unique opportunities in searching for new physics through its high center of mass energy and coherent interactions of large nuclei.  We examine how light weakly interacting vector bosons from a variety of models can be discovered or constrained, over significant parts of their parameter space, through clean displaced vertex signals at the EIC.  Our results indicate that the searches we propose favorably compare with or surpass existing experimental projections for the models examined.  The reach for the new physics that we consider can be markedly improved if ``far backward" particle identification capabilities are included in the   EIC detector complex.

\end{abstract}

\maketitle

\section{Introduction\label{sec:intro}}

A number of experimental observations, as well as conceptual puzzles, lead us to the unavoidable conclusion that new physics beyond the Standard Model (SM) is required for a complete fundamental description of natural phenomena.  For long,  conventional thinking largely assigned new physics to ever shorter distances, corresponding to increasingly larger  energy scales.  However, recent years have seen a surge of interest in searching for new low mass particles that may have evaded discovery due to their feeble interactions with known states.  This view of physics calls for new approaches in both theory and experiment in order to explore various  possibilities.  

It may at first seem that one could always efficiently hide new light states by suppressing their couplings.  Yet, over certain ranges of parameters, the reduced production rates can be compensated for by a concomitant suppression of back ground events, due to the longevity of the hypothesized  particles.  An extreme version of this effect is the basis for the impressive reach of ``beam-dump" experiments, where large amounts of shielding material remove much of the SM  background, while allowing long-lived feebly interacting states to reach a downstream detector.  Less extreme  longevity of the new states has also been used in conventional collider searches, where the macroscopic decay length of a  hypothetical particle would lead to displaced vertices that could greatly reduce background contamination of such signals. 

In this work, we consider models where a light boson can be produced and identified, through the aforementioned displaced decay vertices, at the future Electron-Ion Collider (EIC).  In particular, we will focus on light hidden vector models, characterized by mass scales of $\ord{\rm 100~MeV}$, that are produced by coupling to the electron beam.  We study three such representative models: a standard dark photon, a gauge boson coupled to the $B-L$ symmetry of the SM, and leptophilic gauge bosons which couple to lepton flavor number.

Although similar searches may be conducted at other facilities, such as typical fixed target experiments, the EIC has certain advantages that allow it to provide quite competitive and often complementary probes of new physics.  While some experiments offer significant luminosities, they typically have small center of mass energy $\sqrt{s}$, limiting their reach.  In this respect, the EIC provides a marked strength, given its relatively large beam energies of 18 GeV for the electron and 110 GeV per nucleon for the ion; for a gold nucleus, in the nuclear rest frame this is equivalent to roughly 4 TeV of energy for the electron \cite{Davoudiasl:2021mjy}.  At the same time, the heavy ion beam at the EIC, like in fixed target experiments, can lead to significantly enhanced interaction rates through coherent scattering from large nuclei.  An expected integrated luminosity of $\mathcal{L} = 100$ fb${}^{-1}/ A \approx 0.5$ fb${}^{-1}$ operating in electron-gold collisions \cite{AbdulKhalek:2021gbh}, combined with large beam energy and enhanced interaction rates, can give significant reach in searches for new physics.


\vskip1cm

Some aspects of this work follow from our previous work on phenomenology of axion-like particles at the EIC \cite{Davoudiasl:2021mjy}.  For a similar search for displaced hidden vectors at the MUonE experiment, see \cite{GrillidiCortona:2022kbq,Galon:2022xcl}.  Other recent studies of potential searches for beyond SM physics at the EIC are \cite{Gonderinger:2010yn, Cirigliano:2021img, Liu:2021lan, Yan:2021htf, Li:2021uww, Batell:2022ogj,Zhang:2022zuz,Yan:2022npz}. 

\section{Displaced signal\label{sec:signal}}

\subsection{Vector boson production}

\begin{figure}
    \centering
    \includegraphics[width = \linewidth]{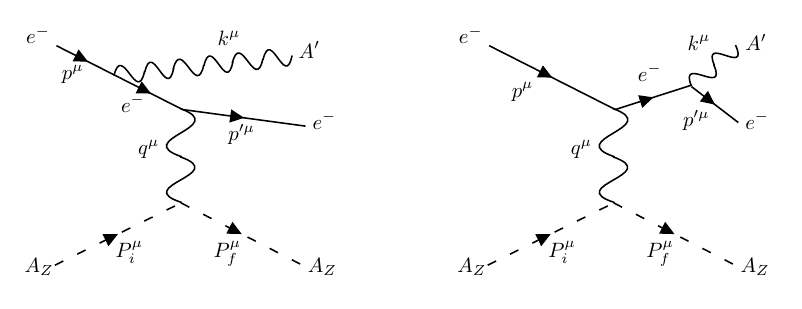}
    \caption{Vector boson production at the EIC.}
    \label{fig:diagram}
\end{figure}

The process under consideration is ultraperipheral production of a massive light vector boson $A'$ at the EIC, represented by the process $e^-A_Z \rightarrow e^-A_Z A'$. If $A'$ is light enough, it is likely that its decay will be displaced relative to the production vertex, which yields a signal with virtually zero SM background. In particular, we will focus on the scenario in which the $A'$ decays into an $e^+e^-$ pair, taking advantage of the EIC's electron tracking capabilities.  We only consider emission of the $A'$ from the electron, and not from the ion; emission from the ion will be suppressed by the nuclear form factor except for fairly light $A'$ masses $m_{A'}$, where other experiments already provide stringent limits.

The Feynman diagrams representing the process are shown in Fig.~\ref{fig:diagram}. Following  Ref.~\cite{Liu:2017htz}, we label the incoming momenta $p^\mu$ for the electron and $P_i^\mu$ for the incoming ion, and the outgoing momenta $p'^\mu$ for the electron, $k$ for the vector boson, and $P_f^\mu$ for the outgoing ion. It is useful to define the virtual photon momentum transfer $q^\mu \equiv P_f^\mu - P_i^\mu$ as well as the four momentum sum $P^\mu \equiv P_f^\mu + P_i^\mu$. Treating the ion as a scalar, the photon-ion interaction vertex is given by
\beq
    iV^\mu(q^2, P_i, P_f) = ieZF(q^2)P^\mu
\eeq
where $F(q^2)$ is the elastic form factor of the nucleus. In this work, we will consider collisions with a gold ($Z=79$, mass number $A=197$), corresponding to a mass of $M=183$ GeV. For the form factor, we  use an approximation of the Fourier transform of the Woods-Saxon distribution applied to a gold nucleus \cite{Klein:1999qj}, given by
\beq \label{eq:form_factor}
    F(q^2) = \frac{3}{q^3R_A^3}\left(\sin{qR_A} - qR_A\cos{qR_A}\right)\frac{1}{1 + a_0^2q^2}
\eeq
where $a_0 = 0.79~{\rm fm}$ and $R_A = (1.1~{\rm fm})A^{1/3}$. 

With these specifications, the amplitude of the Feynman diagrams in Fig.~\ref{fig:diagram} can be computed. Following the EIC Yellow Report \cite{AbdulKhalek:2021gbh}, we take the initial-state lab-frame energies of the electron and ion to be $18\,$GeV and $110\,$GeV/nucleon, respectively. Then, we integrate the squared amplitude over the phase space to obtain the total cross-section. The EIC has electron detection capabilities for any electron within $|\eta| < 3.5$, so when integrating over the phase space, we impose this restriction on the $A'$. The details of the amplitude calculation and cross-section integration are shown in Appendix~\ref{app:signal}.

\subsection{Displacement within the Detector}
If the lifetime of the $A'$ is $\tau$ and it has boost $\gamma_k$ and speed $v_k$ in the lab frame, the probability of it undergoing a displaced decay which is detectable is
\begin{align}
    P_{\rm disp} = e^{-d_{\rm min}/(\gamma_k v_k\tau)} - e^{-d_{\rm max}/(\gamma_k v_k\tau)} \label{eq:prob}
\end{align}
where $d_{\rm min}$ is set by the resolution of the detector, and $d_{\rm max}$ is set  by the geometry of the detector. The lifetime $\tau$ in Eq.~\ref{eq:prob} is given by the inverse width of the $A'$, which depends on the various charges of SM particles under $U(1)_{A'}$; see Appendix~\ref{app:signal} for details. 

To determine $d_{\rm min}$, we refer to the design document for the ECCE detector (now the ePIC collaboration) at the EIC, \cite{Adkins:2022jfp} which provides resolutions for the 2D distance of closest approach (${\rm DCA_{2D}}$) of pions. In particular, ${\rm DCA_{2D}^{\rm min}} < 100{\rm \mu m}$ for almost all track transverse momenta ($p_T$) and pseudorapidities. If we adopt this as the resolution for the ${\rm DCA_{2D}}$ of electrons at the EIC, we can relate $d_{\rm min}$ to the lifetime $\tau$ of the $A'$. In particular, the transverse DCA is defined as the spatial separation between the primary vertex and reconstructed particle paths projected onto the transverse plane (for details, see Fig. \ref{fig:dca} in Appendix~\ref{app:dca} and the following discussion). Assuming $m_e \ll m_{A'}$, we find $d_{\rm min} \approx \gamma_k( {\rm DCA}_{\rm 2D}^{\rm min})/v_k\cos{\theta_k^{\rm lab}}$, where $\theta_k^{\rm lab}$ is the angle the dark boson makes with the electron beam axis in the lab frame. We have chosen $d_{\rm  max} = 1~{\rm m}$, which is in line with the proposed geometry of the ECCE/ePIC detector \cite{Adkins:2022jfp}.

One of the main limiting factors of our production cross-section is the production of dark photons with far-backward pseudorapidities ($\eta < -3.5$), which are beyond the EIC’s current tracking capabilities. Notably, the ECCE detector proposal includes a far-forward detector, the B0 spectrometer, with the capability to track particles with $4 <\eta < 6$\cite{Adkins:2022jfp}.  Thus, we consider a scenario in which a similar detector is installed in the backward region at around $z = -5~{\rm m}$, with the ability to track electrons with $-6 < \eta < -4$. We assume that this detector has a weaker DCA resolution than the rest of the detector, ${\rm {DCA}_{2D}^{min}} = 200\,{\mu\rm m}$, so we take $d_{\rm min} = \gamma_k(200\,{\mu\rm m})/v_k\cos{\theta_k^{\rm lab}}$ and $d_{\rm max} = 5\,{\rm m}$.

\subsection{Signal selection}
The selection criteria we choose to search for displaced vectors at the EIC is the identification of an electron and positron each displaced from the primary vertex. The corresponding cross-section for signal events is given by
\begin{align}
    \sigma_{\rm sig}(g_{A'}) = \int{P_{\rm disp}\frac{d\sigma}{d\gamma_k\,d\eta_k}{d\gamma_k\,d\eta_k}}\,{\cal B}(A'\rightarrow e^+ e^-)
\end{align}
where $\gamma_k$ and $\eta_k$ are the boost and pseudorapidity of the vector in the lab frame, and ${\cal B}(A'\rightarrow e^+ e^-) = \Gamma_{\bar{e}e}/\Gamma$ is the branching fraction into electron-positrion pairs. For the base EIC limits, we numerically integrate $\eta_k$ over the region $|\eta_k| < 3.5$ and $\gamma_k$ from 1 to $(18\,{\rm GeV})/m_{A'}$. For the far-backward detector scenario, we integrate $\eta_k$ from $-4$ to $-6$. The dependence of the RHS on the vector boson-electron coupling $g_{A'}$ comes from the lifetime $\tau$ in $P_{\rm disp}$ and the differential cross-section. Given the size of the displacements considered in this study ($d_{\rm min} \gg 0.1{\rm mm}$), we assume that there is negligible SM background. 

For dark photons in particular, a potentially concerning background is ordinary photon conversion.  Because our signal is concentrated at large $|\eta|$ in the direction of the electron beam, the vast majority of signal events will occur in a region of the proposed ECCE/ePIC detector which is very sparse, consisting of isolated silicon-tracker disks with a separation of $\sim 25$ cm \cite{Adkins:2022jfp}.  As a result, cutting away displaced vertices which originate at the disk could be an effective way to remove photon conversions without losing many signal events.  Reconstruction of the invariant mass of lepton pairs could be another experimental handle to distinguish $A'$ events from photon conversions in order to satisfy our assumption of negligible SM background.

Another potential source of background is misidentification of charged pions, which will be copiously produced in ion collisions, as $e^{\pm}$.  However, in the electron endcap where our signal is concentrated, the fake rate is quite low, approximately $10^{-4}$ \cite{Adkins:2022jfp}.  Since our signal requires both $e^{-}$ and $e^{+}$ as well as the displaced vertex, this background should be negligible.

Finally, there is also the possibility of signal reduction if electrons or positrons from the $A'$ decay are lost down the beam pipe.  We have used our kinematic distributions with some simplifying assumptions to estimate that, conservatively, this rate is no larger than 20\%-30\% even with the signal strongly collimated in the backward direction.  Since our estimate is somewhat crude (and would not affect our projections significantly) we do not include it in our projections, but a future study with a full Monte Carlo detector simulation could take this possibility into account for more accurate bounds.

To place limits on the coupling $g_{A'}$, we find the values of $g_{A'}$ for which
\begin{align}
    {\cal L}\sigma(g_{A'}) \geq n_{\rm max}
\end{align}
where $n_{\rm max} = 3.09$ is the upper limit of the 95\% confidence interval on the mean number of signal events given zero expected background events \cite{Feldman:1997qc}, and we take ${\cal L} = 100\,{\rm fb}^{-1}/A \approx 0.5\,{\rm fb}^{-1}$ in line with Ref.~\cite{AbdulKhalek:2021gbh}.

\section{Model limits\label{sec:limits}}

\begin{figure}
    \centering
    \includegraphics[width=\linewidth]{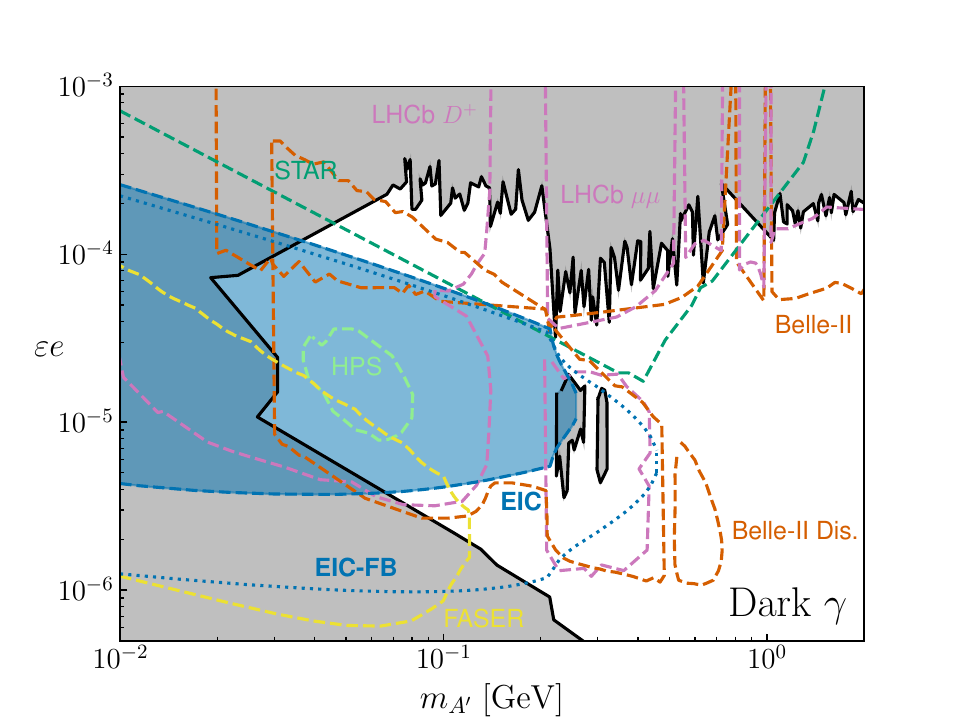}
    \caption{Projected constraints (95\% C.L.) on the kinetic mixing of a dark photon at the EIC (blue filled region).  Exclusion limits (grey filled) include U70/NuCal~\cite{Blumlein:2011mv,Blumlein:2013cua}, Orsay~\cite{Davier:1989wz}, E137~\cite{Bjorken:1988as}, E141~\cite{Riordan:1987aw}, E774~\cite{Bross:1989mp}, (95\% C.L.) and NA48~\cite{NA482:2015wmo}, BaBar~\cite{BaBar:2009lbr,BaBar:2014zli}, KLOE~\cite{KLOE-2:2011hhj,KLOE-2:2012lii,KLOE-2:2016ydq,Anastasi:2015qla}, LHCb\cite{LHCb:2017trq} (90\% C.L.).  Compared to Ref.~\cite{Bauer:2018onh}, we have updated the excluded region to include NA64~\cite{Banerjee:2019pds} at 90\% C.L.  Projections from current experiments (dashed lines) are shown from STAR~\cite{Xu:2022qme}, Belle-II~\cite{Belle-II:2010dht} (90\% C.L.), a second Belle-II projection for displaced decays \cite{Ferber:2022ewf} (90\% C.L., 50 ab$^{-1}$), HPS \cite{Baltzell:2022rpd}, FASER\cite{FASER:2018eoc}, and LHCb\cite{Ilten:2015hya,Ilten:2016tkc,LHCb:2017trq} (95\% C.L).  For an alternative set of Belle-II displaced projections, see \cite{Bandyopadhyay:2022klg}.  To make our plot easier to read, we show projections only for currently operating experiments; for a more complete set of projected bounds, see the Snowmass whitepaper \cite{Batell:2022dpx}.  Finally, we show a projection for inclusion of a ``far-backward'' detector at the EIC (dotted blue line, 95\% C.L.), as described in Sec.~\ref{sec:signal}.  {Digitized plot data are available as ancillary files on arXiv.org.}}
    \label{fig:limit_dark_A}
\end{figure}

\begin{figure}
    \centering
    \includegraphics[width=\linewidth]{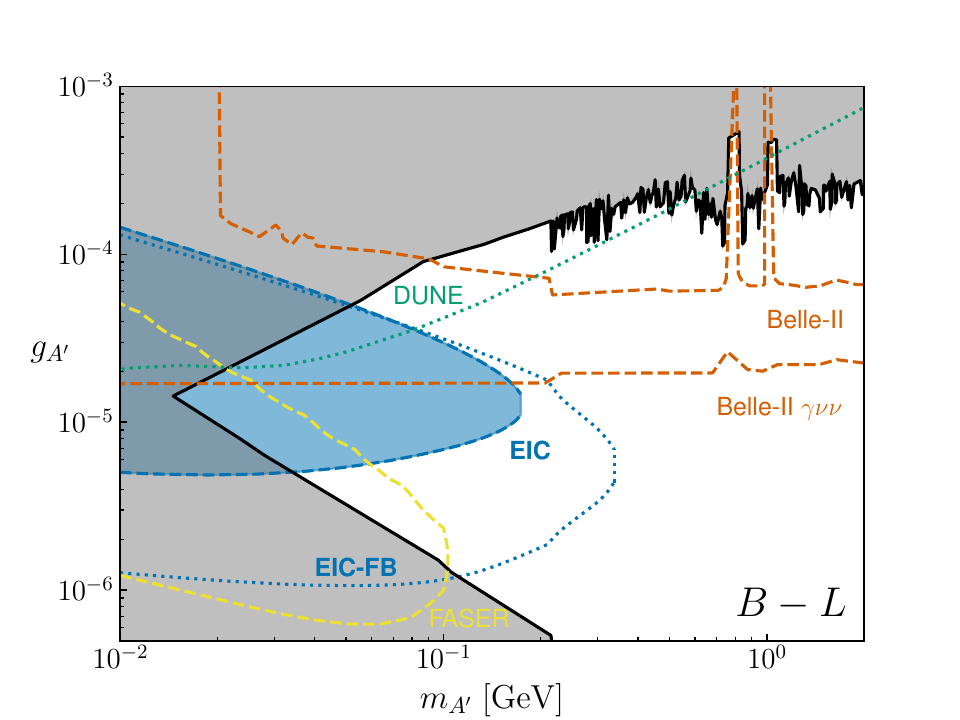}
    \caption{Projected constraints (95\% C.L.) on the interaction strength of a $U(1)_{B-L}$ vector gauge boson at the EIC.  Colors and references follow Fig.~\ref{fig:limit_dark_A}.  The excluded region (grey filled) has been expanded to include constraints from Borexino\cite{Harnik:2012ni} and Texono\cite{Lindner:2018kjo}.  We show additional projections for Belle-II\cite{Belle-II:2018jsg,Bauer:2018onh}, a projection from DUNE \cite{Chakraborty:2021apc}, and a recast FASER projection from DarkCast\cite{Ilten:2018crw,Baruch:2022esd} (\url{https://gitlab.com/darkcast}).}
    \label{fig:limit_B_L}
\end{figure}

\begin{figure}
    \centering
    \includegraphics[width=\linewidth]{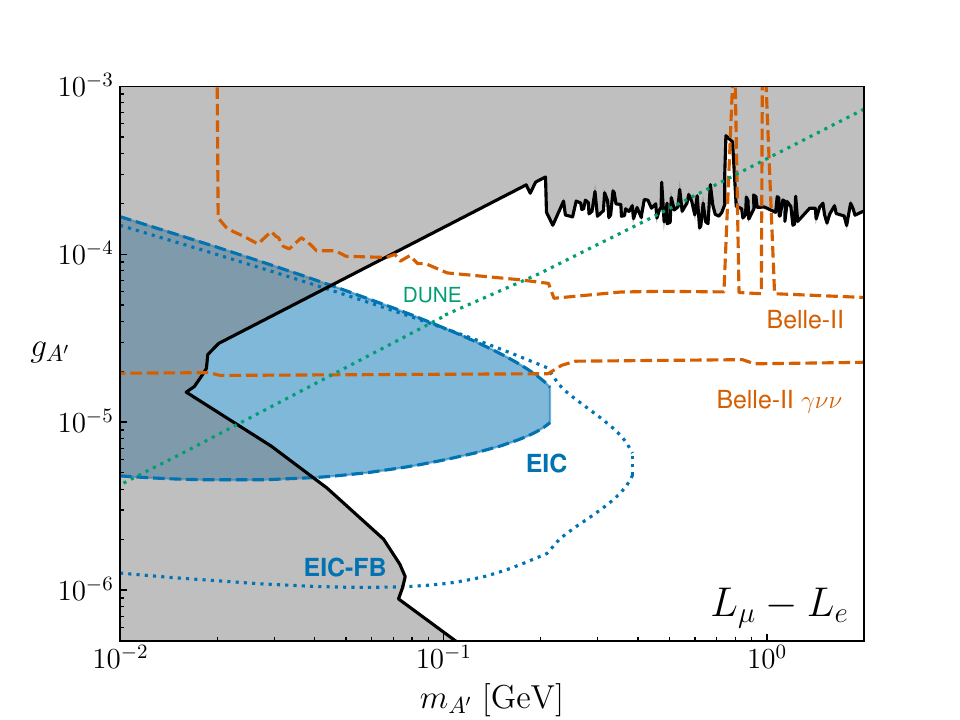}
    \caption{Projected constraints (95\% C.L.) on the interaction strength of a $U(1)_{L_{\mu}-L_e}$ vector gauge boson at the EIC.  Details of the regions shown follow Fig.~\ref{fig:limit_B_L}.  For DUNE, we show a combination of the projections from Ref.~\cite{Chakraborty:2021apc} and from Refs.~\cite{Bauer:2018onh,Wise:2018rnb}.}
    \label{fig:limit_mu_e}
\end{figure}

\begin{figure}
    \centering
    \includegraphics[width=\linewidth]{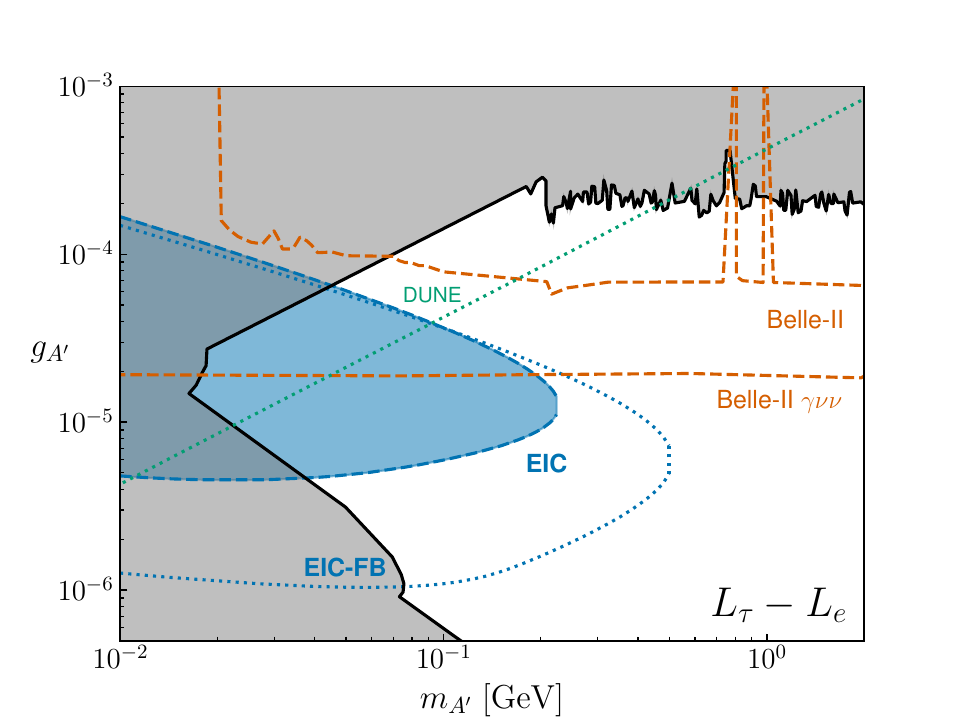}
    \caption{Projected constraints (95\% C.L.) on the interaction strength of a $U(1)_{L_{\tau} - L_e}$ vector gauge boson at the EIC.  Details of the regions shown follow Fig.~\ref{fig:limit_mu_e}.}
    \label{fig:limit_tau_e}
\end{figure}

\subsection{Dark photon}

The dark photon refers to the vector boson of a hidden $U(1)_d$ gauge symmetry under which none of the SM matter fields are charged.  However, one can write down a kinetic mixing term of the form
\beq
\frac{\eps}{2 \cos\theta_W} F_{d\mu \nu}F_Y^{\mu\nu}\,,
\label{kin-mix}
\eeq
between the $U(1)_d$ and the SM hypercharge field strength tensors, denoted by $F_{d\mu \nu}$ and $F_Y^{\mu\nu}$, respectively.  Here, $\eps$ is a small parameter and $\theta_W$ denotes the weak mixing angle.  With the above normalization, after kinetic term diagonalization, SM states of electric charge $Q_e$ couple to the dark photon ($A'$) with strength $Q_e \eps e$, where $e = \sqrt{4 \pi \alpha}$ is the electromagnetic coupling constant; $\alpha \approx 1/137$.  If there are heavy states charged under both gauge sectors, a small value  $\eps \sim e g_d/(16 \pi^2)$ (assuming unit charges) can naturally be induced for the mixing parameter through a one-loop diagram \cite{Holdom:1985ag}, where $g_d$ is the $U(1)_d$ gauge coupling, which can be $\ord{1}$ or less.  The dark photon has been of interest in recent years, in particular as a  connection between a new sector containing DM and the SM \cite{Pospelov:2007mp,Arkani-Hamed:2008hhe}.  For the dark photon, we take $g_{A'} = \varepsilon e$.

In Fig.~\ref{fig:limit_dark_A}, we show our projected limits for displaced dark photons from the EIC.  We also show for comparison various existing limits (grey shaded region) and projected limits from current experiments (dashed lines) and future proposed experiments (dotted lines).  The solid filled region shows our baseline projection using the EPIC detector and a luminosity for gold ion collisions of $\mathcal{L} = 100$ fb$^{-1} / A$.  We see that the EIC can provide significant new constraints on the parameter space for masses $m_{A'} \sim 100$ MeV and moderately weak couplings $\eps e \sim 10^{-5}$.  Even comparing to other projected experimental bounds, the EIC provides useful reach in this parameter space.  The dashed line labelled ``EIC-FB'' shows how the projected bounds from the EIC could be improved by the addition of a ``far-backward'' detector in the direction of the electron beam, as described above in Sec.~\ref{sec:signal}.

\subsection{$B-L$ gauge boson}

In the SM, $B-L$ -- where $B$ and $L$ are the baryon and lepton numbers, respectively -- is not violated by any interaction.  With the addition of three right-handed complete singlet neutrinos, which can lead to non-zero masses for active neutrinos as required by experiment, one can make this quantum number anomaly free.  Hence, it is well-motivated to consider a $U(1)_{B-L}$ gauge symmetry.  In fact, this gauge group was considered early on as an extension of the SM electroweak symmetry \cite{Mohapatra:1974hk,Mohapatra:1974gc,Senjanovic:1975rk}.   Since SM fermions couple with $\ord{1}$ charges (assuming conventional charge assignments) to the corresponding gauge field $A'$, a low mass $m_{A'}\lsim$~GeV for this new vector boson can be phenomenologically viable only for tiny gauge coupling constants $g_{A'}\ll 1$.

Figure \ref{fig:limit_B_L} shows our projected limits on the $B-L$ gauge boson from the EIC, as described for the dark photon search above.  Here again, we see reach for the EIC which is significantly beyond existing bounds and complementary to other planned experimental searches, particularly in the far-backward detector scenario.

\subsection{Leptophilic gauge bosons}

Differences $L_i - L_j$, with  $i,j=e,\mu,\tau$ and $i\neq j$, of lepton flavor numbers are anomaly free in the SM.  One may gauge one of these quantum numbers at a time, and obtain an anomaly free $U(1)_{ij}$ \cite{He:1990pn}.  Like in the case of the $B-L$ gauge interaction, since the SM leptons directly couple to the corresponding gauge field $A'$ with unit charge, the coupling constant here needs to be quite small, $g_{A'}\ll 1$, if the gauge vector boson has a mass $\lsim $~GeV.  

Searches for displaced bosons at the EIC are sensitive to both $(L_\mu - L_e)$ and $(L_\tau - L_e)$ gauge bosons; the third combination $(L_\tau - L_\mu)$ does not couple to electrons and so cannot easily be produced at the EIC.  The limits obtained on each of these hypothetical gauge bosons are shown in Figs.~\ref{fig:limit_mu_e} and \ref{fig:limit_tau_e} in blue.  

See also Ref.~\cite{Moroi:2022qwz} for projected limits from beam-dump experiments at future lepton colliders such as the ILC; we do not include their projections in our plots because they are generally applicable at much weaker $g_{A'}$.

Figures \ref{fig:limit_mu_e} and \ref{fig:limit_tau_e} show our projected limits on leptophilic gauge bosons which couple to electrons, again as described for the dark photon search above.  Qualitatively, these projections are similar to those obtained for the $B-L$ gauge boson scenario and show distinct sensitivity to parts of the parameter space, more strongly in the presence of far-backward detector.

\section{Concluding Remarks\label{sec:conclusions}}

In this work, we have considered how the future EIC capabilities can be used to probe a number of models that include light hidden vector bosons with  masses in the range $\sim \text{few}\times (0.01-0.1)$~GeV and couplings $\ll 1$ to the SM fields.  For all of the new-particle searches we consider, electron-gold ion collisions at the EIC have considerable projected reach for unexplored values of the coupling strength. Compared to other collider experiments, the reach of the EIC is improved due to the enhancement of the cross-section by the nuclear charge $Z$.  Relative to fixed-target experiments, the EIC provides a controlled environment with good detector coverage and a large center-of-mass energy (in the ion rest frame, the electron energy is approximately 4 TeV \cite{Davoudiasl:2021mjy}.)   

We focused on a clean and essentially background-free regime of parameters where the produced vector bosons have sufficiently long lifetimes that lead to displaced decay vertices in the detector.  We found that with realistic integrated luminosity allocations, of order 100~fb$^{-1}/A$, coupling parameters down to $\sim 5\times 10^{-6}$ can be probed.  The search for the low mass physics considered in our work can be significantly improved with the addition of ``far backward" particle detection capabilities at the EIC.  Our estimates suggest that such an addition can push the reach down to couplings $\sim 10^{-6}$ and up to masses $m_{A'} \sim 500$ MeV.

The analysis in this paper mainly assumed the expected excellent electron final state identification capabilities of the future EIC detector.   However, significant branching fraction into $\mu^+ \mu^-$ is present in parts of our model parameter space, particularly at heavier $A'$ masses.  Muon detection capability at the EIC could provide additional statistics and strengthen our bounds somewhat.  In general, muon detection could provide an additional handle on searches for BSM particles more broadly, particularly for heavy particles above the dimuon kinematic limit or final states with more significant SM backgrounds.

The EIC will also possess the capability to collide polarized electrons with ions, which is not a common feature of most current and future experiments that may probe the same mass and coupling parameter space.  In principle, this can give an additional handle to probe the chiral structure of the $A'$ couplings, for example to distinguish a generic dark photon from something like a dark $Z$ boson \cite{Davoudiasl:2012ag}.  We did not examine the prospects for using the EIC electron beam polarization in our study.  However, an analysis using this handle on the new physics would be interesting, which we leave for future work.


\section*{Acknowledgements}
We thank G. Grilli di Cortona for helpful communication regarding Ref.~\cite{GrillidiCortona:2022kbq}.   We also thank M. Williams for comments on an earlier version of the paper. 
 This work is supported by the U.S. Department of Energy under Grant Contracts DE-SC0012704 (H.~D.) and DE-SC0010005 (E.~N. and R.~M.).

\begin{raggedright}
\bibliography{eic-displaced}

\begin{thebibliography}{57}
\expandafter\ifx\csname natexlab\endcsname\relax\def\natexlab#1{#1}\fi
\expandafter\ifx\csname bibnamefont\endcsname\relax
  \def\bibnamefont#1{#1}\fi
\expandafter\ifx\csname bibfnamefont\endcsname\relax
  \def\bibfnamefont#1{#1}\fi
\expandafter\ifx\csname citenamefont\endcsname\relax
  \def\citenamefont#1{#1}\fi
\expandafter\ifx\csname url\endcsname\relax
  \def\url#1{\texttt{#1}}\fi
\expandafter\ifx\csname urlprefix\endcsname\relax\def\urlprefix{URL }\fi
\providecommand{\bibinfo}[2]{#2}
\providecommand{\eprint}[2][]{\url{#2}}

\bibitem[{\citenamefont{Davoudiasl et~al.}(2023)\citenamefont{Davoudiasl,
  Marcarelli, and Neil}}]{Davoudiasl:2021mjy}
\bibinfo{author}{\bibfnamefont{H.}~\bibnamefont{Davoudiasl}},
  \bibinfo{author}{\bibfnamefont{R.}~\bibnamefont{Marcarelli}},
  \bibnamefont{and} \bibinfo{author}{\bibfnamefont{E.~T.} \bibnamefont{Neil}},
  \bibinfo{journal}{JHEP} \textbf{\bibinfo{volume}{02}}, \bibinfo{pages}{071}
  (\bibinfo{year}{2023}), \eprint{2112.04513}.

\bibitem[{\citenamefont{Abdul~Khalek et~al.}(2022)}]{AbdulKhalek:2021gbh}
\bibinfo{author}{\bibfnamefont{R.}~\bibnamefont{Abdul~Khalek}}
  \bibnamefont{et~al.}, \bibinfo{journal}{Nucl. Phys. A}
  \textbf{\bibinfo{volume}{1026}}, \bibinfo{pages}{122447}
  (\bibinfo{year}{2022}), \eprint{2103.05419}.

\bibitem[{\citenamefont{Grilli~di Cortona and
  Nardi}(2022)}]{GrillidiCortona:2022kbq}
\bibinfo{author}{\bibfnamefont{G.}~\bibnamefont{Grilli~di Cortona}}
  \bibnamefont{and} \bibinfo{author}{\bibfnamefont{E.}~\bibnamefont{Nardi}},
  \bibinfo{journal}{Phys. Rev. D} \textbf{\bibinfo{volume}{105}},
  \bibinfo{pages}{L111701} (\bibinfo{year}{2022}), \eprint{2204.04227}.

\bibitem[{\citenamefont{Galon et~al.}(2023)\citenamefont{Galon, Shih, and
  Wang}}]{Galon:2022xcl}
\bibinfo{author}{\bibfnamefont{I.}~\bibnamefont{Galon}},
  \bibinfo{author}{\bibfnamefont{D.}~\bibnamefont{Shih}}, \bibnamefont{and}
  \bibinfo{author}{\bibfnamefont{I.~R.} \bibnamefont{Wang}},
  \bibinfo{journal}{Phys. Rev. D} \textbf{\bibinfo{volume}{107}},
  \bibinfo{pages}{095003} (\bibinfo{year}{2023}), \eprint{2202.08843}.

\bibitem[{\citenamefont{Gonderinger and
  Ramsey-Musolf}(2010)}]{Gonderinger:2010yn}
\bibinfo{author}{\bibfnamefont{M.}~\bibnamefont{Gonderinger}} \bibnamefont{and}
  \bibinfo{author}{\bibfnamefont{M.~J.} \bibnamefont{Ramsey-Musolf}},
  \bibinfo{journal}{JHEP} \textbf{\bibinfo{volume}{11}}, \bibinfo{pages}{045}
  (\bibinfo{year}{2010}), \bibinfo{note}{[Erratum: JHEP 05, 047 (2012)]},
  \eprint{1006.5063}.

\bibitem[{\citenamefont{Cirigliano et~al.}(2021)\citenamefont{Cirigliano,
  Fuyuto, Lee, Mereghetti, and Yan}}]{Cirigliano:2021img}
\bibinfo{author}{\bibfnamefont{V.}~\bibnamefont{Cirigliano}},
  \bibinfo{author}{\bibfnamefont{K.}~\bibnamefont{Fuyuto}},
  \bibinfo{author}{\bibfnamefont{C.}~\bibnamefont{Lee}},
  \bibinfo{author}{\bibfnamefont{E.}~\bibnamefont{Mereghetti}},
  \bibnamefont{and} \bibinfo{author}{\bibfnamefont{B.}~\bibnamefont{Yan}},
  \bibinfo{journal}{JHEP} \textbf{\bibinfo{volume}{03}}, \bibinfo{pages}{256}
  (\bibinfo{year}{2021}), \eprint{2102.06176}.

\bibitem[{\citenamefont{Liu and Yan}(2023)}]{Liu:2021lan}
\bibinfo{author}{\bibfnamefont{Y.}~\bibnamefont{Liu}} \bibnamefont{and}
  \bibinfo{author}{\bibfnamefont{B.}~\bibnamefont{Yan}},
  \bibinfo{journal}{Chin. Phys. C} \textbf{\bibinfo{volume}{47}},
  \bibinfo{pages}{043113} (\bibinfo{year}{2023}), \eprint{2112.02477}.

\bibitem[{\citenamefont{Yan et~al.}(2021)\citenamefont{Yan, Yu, and
  Yuan}}]{Yan:2021htf}
\bibinfo{author}{\bibfnamefont{B.}~\bibnamefont{Yan}},
  \bibinfo{author}{\bibfnamefont{Z.}~\bibnamefont{Yu}}, \bibnamefont{and}
  \bibinfo{author}{\bibfnamefont{C.~P.} \bibnamefont{Yuan}},
  \bibinfo{journal}{Phys. Lett. B} \textbf{\bibinfo{volume}{822}},
  \bibinfo{pages}{136697} (\bibinfo{year}{2021}), \eprint{2107.02134}.

\bibitem[{\citenamefont{Li et~al.}(2022)\citenamefont{Li, Yan, and
  Yuan}}]{Li:2021uww}
\bibinfo{author}{\bibfnamefont{H.~T.} \bibnamefont{Li}},
  \bibinfo{author}{\bibfnamefont{B.}~\bibnamefont{Yan}}, \bibnamefont{and}
  \bibinfo{author}{\bibfnamefont{C.~P.} \bibnamefont{Yuan}},
  \bibinfo{journal}{Phys. Lett. B} \textbf{\bibinfo{volume}{833}},
  \bibinfo{pages}{137300} (\bibinfo{year}{2022}), \eprint{2112.07747}.

\bibitem[{\citenamefont{Batell et~al.}(2023)\citenamefont{Batell, Ghosh, Han,
  and Xie}}]{Batell:2022ogj}
\bibinfo{author}{\bibfnamefont{B.}~\bibnamefont{Batell}},
  \bibinfo{author}{\bibfnamefont{T.}~\bibnamefont{Ghosh}},
  \bibinfo{author}{\bibfnamefont{T.}~\bibnamefont{Han}}, \bibnamefont{and}
  \bibinfo{author}{\bibfnamefont{K.}~\bibnamefont{Xie}},
  \bibinfo{journal}{JHEP} \textbf{\bibinfo{volume}{03}}, \bibinfo{pages}{020}
  (\bibinfo{year}{2023}), \eprint{2210.09287}.

\bibitem[{\citenamefont{Zhang et~al.}(2023)}]{Zhang:2022zuz}
\bibinfo{author}{\bibfnamefont{J.~L.} \bibnamefont{Zhang}}
  \bibnamefont{et~al.}, \bibinfo{journal}{Nucl. Instrum. Meth. A}
  \textbf{\bibinfo{volume}{1053}}, \bibinfo{pages}{168276}
  (\bibinfo{year}{2023}), \eprint{2207.10261}.

\bibitem[{\citenamefont{Yan}(2022)}]{Yan:2022npz}
\bibinfo{author}{\bibfnamefont{B.}~\bibnamefont{Yan}}, \bibinfo{journal}{Phys.
  Lett. B} \textbf{\bibinfo{volume}{833}}, \bibinfo{pages}{137384}
  (\bibinfo{year}{2022}), \eprint{2203.01510}.

\bibitem[{\citenamefont{Liu and Miller}(2017)}]{Liu:2017htz}
\bibinfo{author}{\bibfnamefont{Y.-S.} \bibnamefont{Liu}} \bibnamefont{and}
  \bibinfo{author}{\bibfnamefont{G.~A.} \bibnamefont{Miller}},
  \bibinfo{journal}{Phys. Rev. D} \textbf{\bibinfo{volume}{96}},
  \bibinfo{pages}{016004} (\bibinfo{year}{2017}), \eprint{1705.01633}.

\bibitem[{\citenamefont{Klein and Nystrand}(1999)}]{Klein:1999qj}
\bibinfo{author}{\bibfnamefont{S.}~\bibnamefont{Klein}} \bibnamefont{and}
  \bibinfo{author}{\bibfnamefont{J.}~\bibnamefont{Nystrand}},
  \bibinfo{journal}{Phys. Rev. C} \textbf{\bibinfo{volume}{60}},
  \bibinfo{pages}{014903} (\bibinfo{year}{1999}), \eprint{hep-ph/9902259}.

\bibitem[{\citenamefont{Adkins et~al.}(2022)}]{Adkins:2022jfp}
\bibinfo{author}{\bibfnamefont{J.~K.} \bibnamefont{Adkins}}
  \bibnamefont{et~al.} (\bibinfo{year}{2022}), \eprint{2209.02580}.

\bibitem[{\citenamefont{Feldman and Cousins}(1998)}]{Feldman:1997qc}
\bibinfo{author}{\bibfnamefont{G.~J.} \bibnamefont{Feldman}} \bibnamefont{and}
  \bibinfo{author}{\bibfnamefont{R.~D.} \bibnamefont{Cousins}},
  \bibinfo{journal}{Phys. Rev. D} \textbf{\bibinfo{volume}{57}},
  \bibinfo{pages}{3873} (\bibinfo{year}{1998}), \eprint{physics/9711021}.

\bibitem[{\citenamefont{Blumlein and Brunner}(2011)}]{Blumlein:2011mv}
\bibinfo{author}{\bibfnamefont{J.}~\bibnamefont{Blumlein}} \bibnamefont{and}
  \bibinfo{author}{\bibfnamefont{J.}~\bibnamefont{Brunner}},
  \bibinfo{journal}{Phys. Lett. B} \textbf{\bibinfo{volume}{701}},
  \bibinfo{pages}{155} (\bibinfo{year}{2011}), \eprint{1104.2747}.

\bibitem[{\citenamefont{Bl\"umlein and Brunner}(2014)}]{Blumlein:2013cua}
\bibinfo{author}{\bibfnamefont{J.}~\bibnamefont{Bl\"umlein}} \bibnamefont{and}
  \bibinfo{author}{\bibfnamefont{J.}~\bibnamefont{Brunner}},
  \bibinfo{journal}{Phys. Lett. B} \textbf{\bibinfo{volume}{731}},
  \bibinfo{pages}{320} (\bibinfo{year}{2014}), \eprint{1311.3870}.

\bibitem[{\citenamefont{Davier and Nguyen~Ngoc}(1989)}]{Davier:1989wz}
\bibinfo{author}{\bibfnamefont{M.}~\bibnamefont{Davier}} \bibnamefont{and}
  \bibinfo{author}{\bibfnamefont{H.}~\bibnamefont{Nguyen~Ngoc}},
  \bibinfo{journal}{Phys. Lett. B} \textbf{\bibinfo{volume}{229}},
  \bibinfo{pages}{150} (\bibinfo{year}{1989}).

\bibitem[{\citenamefont{Bjorken et~al.}(1988)\citenamefont{Bjorken, Ecklund,
  Nelson, Abashian, Church, Lu, Mo, Nunamaker, and Rassmann}}]{Bjorken:1988as}
\bibinfo{author}{\bibfnamefont{J.~D.} \bibnamefont{Bjorken}},
  \bibinfo{author}{\bibfnamefont{S.}~\bibnamefont{Ecklund}},
  \bibinfo{author}{\bibfnamefont{W.~R.} \bibnamefont{Nelson}},
  \bibinfo{author}{\bibfnamefont{A.}~\bibnamefont{Abashian}},
  \bibinfo{author}{\bibfnamefont{C.}~\bibnamefont{Church}},
  \bibinfo{author}{\bibfnamefont{B.}~\bibnamefont{Lu}},
  \bibinfo{author}{\bibfnamefont{L.~W.} \bibnamefont{Mo}},
  \bibinfo{author}{\bibfnamefont{T.~A.} \bibnamefont{Nunamaker}},
  \bibnamefont{and} \bibinfo{author}{\bibfnamefont{P.}~\bibnamefont{Rassmann}},
  \bibinfo{journal}{Phys. Rev. D} \textbf{\bibinfo{volume}{38}},
  \bibinfo{pages}{3375} (\bibinfo{year}{1988}).

\bibitem[{\citenamefont{Riordan et~al.}(1987)}]{Riordan:1987aw}
\bibinfo{author}{\bibfnamefont{E.~M.} \bibnamefont{Riordan}}
  \bibnamefont{et~al.}, \bibinfo{journal}{Phys. Rev. Lett.}
  \textbf{\bibinfo{volume}{59}}, \bibinfo{pages}{755} (\bibinfo{year}{1987}).

\bibitem[{\citenamefont{Bross et~al.}(1991)\citenamefont{Bross, Crisler,
  Pordes, Volk, Errede, and Wrbanek}}]{Bross:1989mp}
\bibinfo{author}{\bibfnamefont{A.}~\bibnamefont{Bross}},
  \bibinfo{author}{\bibfnamefont{M.}~\bibnamefont{Crisler}},
  \bibinfo{author}{\bibfnamefont{S.~H.} \bibnamefont{Pordes}},
  \bibinfo{author}{\bibfnamefont{J.}~\bibnamefont{Volk}},
  \bibinfo{author}{\bibfnamefont{S.}~\bibnamefont{Errede}}, \bibnamefont{and}
  \bibinfo{author}{\bibfnamefont{J.}~\bibnamefont{Wrbanek}},
  \bibinfo{journal}{Phys. Rev. Lett.} \textbf{\bibinfo{volume}{67}},
  \bibinfo{pages}{2942} (\bibinfo{year}{1991}).

\bibitem[{\citenamefont{Batley et~al.}(2015)}]{NA482:2015wmo}
\bibinfo{author}{\bibfnamefont{J.~R.} \bibnamefont{Batley}}
  \bibnamefont{et~al.} (\bibinfo{collaboration}{NA48/2}),
  \bibinfo{journal}{Phys. Lett. B} \textbf{\bibinfo{volume}{746}},
  \bibinfo{pages}{178} (\bibinfo{year}{2015}), \eprint{1504.00607}.

\bibitem[{\citenamefont{Aubert et~al.}(2009)}]{BaBar:2009lbr}
\bibinfo{author}{\bibfnamefont{B.}~\bibnamefont{Aubert}} \bibnamefont{et~al.}
  (\bibinfo{collaboration}{BaBar}), \bibinfo{journal}{Phys. Rev. Lett.}
  \textbf{\bibinfo{volume}{103}}, \bibinfo{pages}{081803}
  (\bibinfo{year}{2009}), \eprint{0905.4539}.

\bibitem[{\citenamefont{Lees et~al.}(2014)}]{BaBar:2014zli}
\bibinfo{author}{\bibfnamefont{J.~P.} \bibnamefont{Lees}} \bibnamefont{et~al.}
  (\bibinfo{collaboration}{BaBar}), \bibinfo{journal}{Phys. Rev. Lett.}
  \textbf{\bibinfo{volume}{113}}, \bibinfo{pages}{201801}
  (\bibinfo{year}{2014}), \eprint{1406.2980}.

\bibitem[{\citenamefont{Archilli et~al.}(2012)}]{KLOE-2:2011hhj}
\bibinfo{author}{\bibfnamefont{F.}~\bibnamefont{Archilli}} \bibnamefont{et~al.}
  (\bibinfo{collaboration}{KLOE-2}), \bibinfo{journal}{Phys. Lett. B}
  \textbf{\bibinfo{volume}{706}}, \bibinfo{pages}{251} (\bibinfo{year}{2012}),
  \eprint{1110.0411}.

\bibitem[{\citenamefont{Babusci et~al.}(2013)}]{KLOE-2:2012lii}
\bibinfo{author}{\bibfnamefont{D.}~\bibnamefont{Babusci}} \bibnamefont{et~al.}
  (\bibinfo{collaboration}{KLOE-2}), \bibinfo{journal}{Phys. Lett. B}
  \textbf{\bibinfo{volume}{720}}, \bibinfo{pages}{111} (\bibinfo{year}{2013}),
  \eprint{1210.3927}.

\bibitem[{\citenamefont{Anastasi et~al.}(2016)}]{KLOE-2:2016ydq}
\bibinfo{author}{\bibfnamefont{A.}~\bibnamefont{Anastasi}} \bibnamefont{et~al.}
  (\bibinfo{collaboration}{KLOE-2}), \bibinfo{journal}{Phys. Lett. B}
  \textbf{\bibinfo{volume}{757}}, \bibinfo{pages}{356} (\bibinfo{year}{2016}),
  \eprint{1603.06086}.

\bibitem[{\citenamefont{Anastasi et~al.}(2015)}]{Anastasi:2015qla}
\bibinfo{author}{\bibfnamefont{A.}~\bibnamefont{Anastasi}}
  \bibnamefont{et~al.}, \bibinfo{journal}{Phys. Lett. B}
  \textbf{\bibinfo{volume}{750}}, \bibinfo{pages}{633} (\bibinfo{year}{2015}),
  \eprint{1509.00740}.

\bibitem[{\citenamefont{Aaij et~al.}(2018)}]{LHCb:2017trq}
\bibinfo{author}{\bibfnamefont{R.}~\bibnamefont{Aaij}} \bibnamefont{et~al.}
  (\bibinfo{collaboration}{LHCb}), \bibinfo{journal}{Phys. Rev. Lett.}
  \textbf{\bibinfo{volume}{120}}, \bibinfo{pages}{061801}
  (\bibinfo{year}{2018}), \eprint{1710.02867}.

\bibitem[{\citenamefont{Bauer et~al.}(2018)\citenamefont{Bauer, Foldenauer, and
  Jaeckel}}]{Bauer:2018onh}
\bibinfo{author}{\bibfnamefont{M.}~\bibnamefont{Bauer}},
  \bibinfo{author}{\bibfnamefont{P.}~\bibnamefont{Foldenauer}},
  \bibnamefont{and} \bibinfo{author}{\bibfnamefont{J.}~\bibnamefont{Jaeckel}},
  \bibinfo{journal}{JHEP} \textbf{\bibinfo{volume}{07}}, \bibinfo{pages}{094}
  (\bibinfo{year}{2018}), \eprint{1803.05466}.

\bibitem[{\citenamefont{Banerjee et~al.}(2019)}]{Banerjee:2019pds}
\bibinfo{author}{\bibfnamefont{D.}~\bibnamefont{Banerjee}}
  \bibnamefont{et~al.}, \bibinfo{journal}{Phys. Rev. Lett.}
  \textbf{\bibinfo{volume}{123}}, \bibinfo{pages}{121801}
  (\bibinfo{year}{2019}), \eprint{1906.00176}.

\bibitem[{\citenamefont{Xu et~al.}(2022)\citenamefont{Xu, Lewis, Wang,
  Brandenburg, and Ruan}}]{Xu:2022qme}
\bibinfo{author}{\bibfnamefont{I.}~\bibnamefont{Xu}},
  \bibinfo{author}{\bibfnamefont{N.}~\bibnamefont{Lewis}},
  \bibinfo{author}{\bibfnamefont{X.}~\bibnamefont{Wang}},
  \bibinfo{author}{\bibfnamefont{J.~D.} \bibnamefont{Brandenburg}},
  \bibnamefont{and} \bibinfo{author}{\bibfnamefont{L.}~\bibnamefont{Ruan}}
  (\bibinfo{year}{2022}), \eprint{2211.02132}.

\bibitem[{\citenamefont{Abe et~al.}(2010)}]{Belle-II:2010dht}
\bibinfo{author}{\bibfnamefont{T.}~\bibnamefont{Abe}} \bibnamefont{et~al.}
  (\bibinfo{collaboration}{Belle-II}) (\bibinfo{year}{2010}),
  \eprint{1011.0352}.

\bibitem[{\citenamefont{Ferber et~al.}(2022)\citenamefont{Ferber, Garcia-Cely,
  and Schmidt-Hoberg}}]{Ferber:2022ewf}
\bibinfo{author}{\bibfnamefont{T.}~\bibnamefont{Ferber}},
  \bibinfo{author}{\bibfnamefont{C.}~\bibnamefont{Garcia-Cely}},
  \bibnamefont{and}
  \bibinfo{author}{\bibfnamefont{K.}~\bibnamefont{Schmidt-Hoberg}},
  \bibinfo{journal}{Phys. Lett. B} \textbf{\bibinfo{volume}{833}},
  \bibinfo{pages}{137373} (\bibinfo{year}{2022}), \eprint{2202.03452}.

\bibitem[{\citenamefont{Baltzell et~al.}(2022)}]{Baltzell:2022rpd}
\bibinfo{author}{\bibfnamefont{N.}~\bibnamefont{Baltzell}} \bibnamefont{et~al.}
  (\bibinfo{year}{2022}), \eprint{2203.08324}.

\bibitem[{\citenamefont{Ariga et~al.}(2019)}]{FASER:2018eoc}
\bibinfo{author}{\bibfnamefont{A.}~\bibnamefont{Ariga}} \bibnamefont{et~al.}
  (\bibinfo{collaboration}{FASER}), \bibinfo{journal}{Phys. Rev. D}
  \textbf{\bibinfo{volume}{99}}, \bibinfo{pages}{095011}
  (\bibinfo{year}{2019}), \eprint{1811.12522}.

\bibitem[{\citenamefont{Ilten et~al.}(2015)\citenamefont{Ilten, Thaler,
  Williams, and Xue}}]{Ilten:2015hya}
\bibinfo{author}{\bibfnamefont{P.}~\bibnamefont{Ilten}},
  \bibinfo{author}{\bibfnamefont{J.}~\bibnamefont{Thaler}},
  \bibinfo{author}{\bibfnamefont{M.}~\bibnamefont{Williams}}, \bibnamefont{and}
  \bibinfo{author}{\bibfnamefont{W.}~\bibnamefont{Xue}},
  \bibinfo{journal}{Phys. Rev. D} \textbf{\bibinfo{volume}{92}},
  \bibinfo{pages}{115017} (\bibinfo{year}{2015}), \eprint{1509.06765}.

\bibitem[{\citenamefont{Ilten et~al.}(2016)\citenamefont{Ilten, Soreq, Thaler,
  Williams, and Xue}}]{Ilten:2016tkc}
\bibinfo{author}{\bibfnamefont{P.}~\bibnamefont{Ilten}},
  \bibinfo{author}{\bibfnamefont{Y.}~\bibnamefont{Soreq}},
  \bibinfo{author}{\bibfnamefont{J.}~\bibnamefont{Thaler}},
  \bibinfo{author}{\bibfnamefont{M.}~\bibnamefont{Williams}}, \bibnamefont{and}
  \bibinfo{author}{\bibfnamefont{W.}~\bibnamefont{Xue}},
  \bibinfo{journal}{Phys. Rev. Lett.} \textbf{\bibinfo{volume}{116}},
  \bibinfo{pages}{251803} (\bibinfo{year}{2016}), \eprint{1603.08926}.

\bibitem[{\citenamefont{Bandyopadhyay et~al.}(2022)\citenamefont{Bandyopadhyay,
  Chakraborty, and Trifinopoulos}}]{Bandyopadhyay:2022klg}
\bibinfo{author}{\bibfnamefont{T.}~\bibnamefont{Bandyopadhyay}},
  \bibinfo{author}{\bibfnamefont{S.}~\bibnamefont{Chakraborty}},
  \bibnamefont{and}
  \bibinfo{author}{\bibfnamefont{S.}~\bibnamefont{Trifinopoulos}},
  \bibinfo{journal}{JHEP} \textbf{\bibinfo{volume}{05}}, \bibinfo{pages}{141}
  (\bibinfo{year}{2022}), \eprint{2203.03280}.

\bibitem[{\citenamefont{Batell et~al.}(2022)\citenamefont{Batell, Blinov,
  Hearty, and McGehee}}]{Batell:2022dpx}
\bibinfo{author}{\bibfnamefont{B.}~\bibnamefont{Batell}},
  \bibinfo{author}{\bibfnamefont{N.}~\bibnamefont{Blinov}},
  \bibinfo{author}{\bibfnamefont{C.}~\bibnamefont{Hearty}}, \bibnamefont{and}
  \bibinfo{author}{\bibfnamefont{R.}~\bibnamefont{McGehee}}, in
  \emph{\bibinfo{booktitle}{{2022 Snowmass Summer Study}}}
  (\bibinfo{year}{2022}), \eprint{2207.06905}.

\bibitem[{\citenamefont{Harnik et~al.}(2012)\citenamefont{Harnik, Kopp, and
  Machado}}]{Harnik:2012ni}
\bibinfo{author}{\bibfnamefont{R.}~\bibnamefont{Harnik}},
  \bibinfo{author}{\bibfnamefont{J.}~\bibnamefont{Kopp}}, \bibnamefont{and}
  \bibinfo{author}{\bibfnamefont{P.~A.~N.} \bibnamefont{Machado}},
  \bibinfo{journal}{JCAP} \textbf{\bibinfo{volume}{07}}, \bibinfo{pages}{026}
  (\bibinfo{year}{2012}), \eprint{1202.6073}.

\bibitem[{\citenamefont{Lindner et~al.}(2018)\citenamefont{Lindner, Queiroz,
  Rodejohann, and Xu}}]{Lindner:2018kjo}
\bibinfo{author}{\bibfnamefont{M.}~\bibnamefont{Lindner}},
  \bibinfo{author}{\bibfnamefont{F.~S.} \bibnamefont{Queiroz}},
  \bibinfo{author}{\bibfnamefont{W.}~\bibnamefont{Rodejohann}},
  \bibnamefont{and} \bibinfo{author}{\bibfnamefont{X.-J.} \bibnamefont{Xu}},
  \bibinfo{journal}{JHEP} \textbf{\bibinfo{volume}{05}}, \bibinfo{pages}{098}
  (\bibinfo{year}{2018}), \eprint{1803.00060}.

\bibitem[{\citenamefont{Altmannshofer et~al.}(2019)}]{Belle-II:2018jsg}
\bibinfo{author}{\bibfnamefont{W.}~\bibnamefont{Altmannshofer}}
  \bibnamefont{et~al.} (\bibinfo{collaboration}{Belle-II}),
  \bibinfo{journal}{PTEP} \textbf{\bibinfo{volume}{2019}},
  \bibinfo{pages}{123C01} (\bibinfo{year}{2019}), \bibinfo{note}{[Erratum: PTEP
  2020, 029201 (2020)]}, \eprint{1808.10567}.

\bibitem[{\citenamefont{Chakraborty et~al.}(2022)\citenamefont{Chakraborty,
  Das, Goswami, and Roy}}]{Chakraborty:2021apc}
\bibinfo{author}{\bibfnamefont{K.}~\bibnamefont{Chakraborty}},
  \bibinfo{author}{\bibfnamefont{A.}~\bibnamefont{Das}},
  \bibinfo{author}{\bibfnamefont{S.}~\bibnamefont{Goswami}}, \bibnamefont{and}
  \bibinfo{author}{\bibfnamefont{S.}~\bibnamefont{Roy}},
  \bibinfo{journal}{JHEP} \textbf{\bibinfo{volume}{04}}, \bibinfo{pages}{008}
  (\bibinfo{year}{2022}), \eprint{2111.08767}.

\bibitem[{\citenamefont{Ilten et~al.}(2018)\citenamefont{Ilten, Soreq,
  Williams, and Xue}}]{Ilten:2018crw}
\bibinfo{author}{\bibfnamefont{P.}~\bibnamefont{Ilten}},
  \bibinfo{author}{\bibfnamefont{Y.}~\bibnamefont{Soreq}},
  \bibinfo{author}{\bibfnamefont{M.}~\bibnamefont{Williams}}, \bibnamefont{and}
  \bibinfo{author}{\bibfnamefont{W.}~\bibnamefont{Xue}},
  \bibinfo{journal}{JHEP} \textbf{\bibinfo{volume}{06}}, \bibinfo{pages}{004}
  (\bibinfo{year}{2018}), \eprint{1801.04847}.

\bibitem[{\citenamefont{Baruch et~al.}(2022)\citenamefont{Baruch, Ilten, Soreq,
  and Williams}}]{Baruch:2022esd}
\bibinfo{author}{\bibfnamefont{C.}~\bibnamefont{Baruch}},
  \bibinfo{author}{\bibfnamefont{P.}~\bibnamefont{Ilten}},
  \bibinfo{author}{\bibfnamefont{Y.}~\bibnamefont{Soreq}}, \bibnamefont{and}
  \bibinfo{author}{\bibfnamefont{M.}~\bibnamefont{Williams}},
  \bibinfo{journal}{JHEP} \textbf{\bibinfo{volume}{11}}, \bibinfo{pages}{124}
  (\bibinfo{year}{2022}), \eprint{2206.08563}.

\bibitem[{\citenamefont{Wise and Zhang}(2018)}]{Wise:2018rnb}
\bibinfo{author}{\bibfnamefont{M.~B.} \bibnamefont{Wise}} \bibnamefont{and}
  \bibinfo{author}{\bibfnamefont{Y.}~\bibnamefont{Zhang}},
  \bibinfo{journal}{JHEP} \textbf{\bibinfo{volume}{06}}, \bibinfo{pages}{053}
  (\bibinfo{year}{2018}), \eprint{1803.00591}.

\bibitem[{\citenamefont{Holdom}(1986)}]{Holdom:1985ag}
\bibinfo{author}{\bibfnamefont{B.}~\bibnamefont{Holdom}},
  \bibinfo{journal}{Phys. Lett. B} \textbf{\bibinfo{volume}{166}},
  \bibinfo{pages}{196} (\bibinfo{year}{1986}).

\bibitem[{\citenamefont{Pospelov et~al.}(2008)\citenamefont{Pospelov, Ritz, and
  Voloshin}}]{Pospelov:2007mp}
\bibinfo{author}{\bibfnamefont{M.}~\bibnamefont{Pospelov}},
  \bibinfo{author}{\bibfnamefont{A.}~\bibnamefont{Ritz}}, \bibnamefont{and}
  \bibinfo{author}{\bibfnamefont{M.~B.} \bibnamefont{Voloshin}},
  \bibinfo{journal}{Phys. Lett. B} \textbf{\bibinfo{volume}{662}},
  \bibinfo{pages}{53} (\bibinfo{year}{2008}), \eprint{0711.4866}.

\bibitem[{\citenamefont{Arkani-Hamed et~al.}(2009)\citenamefont{Arkani-Hamed,
  Finkbeiner, Slatyer, and Weiner}}]{Arkani-Hamed:2008hhe}
\bibinfo{author}{\bibfnamefont{N.}~\bibnamefont{Arkani-Hamed}},
  \bibinfo{author}{\bibfnamefont{D.~P.} \bibnamefont{Finkbeiner}},
  \bibinfo{author}{\bibfnamefont{T.~R.} \bibnamefont{Slatyer}},
  \bibnamefont{and} \bibinfo{author}{\bibfnamefont{N.}~\bibnamefont{Weiner}},
  \bibinfo{journal}{Phys. Rev. D} \textbf{\bibinfo{volume}{79}},
  \bibinfo{pages}{015014} (\bibinfo{year}{2009}), \eprint{0810.0713}.

\bibitem[{\citenamefont{Mohapatra and
  Pati}(1975{\natexlab{a}})}]{Mohapatra:1974hk}
\bibinfo{author}{\bibfnamefont{R.~N.} \bibnamefont{Mohapatra}}
  \bibnamefont{and} \bibinfo{author}{\bibfnamefont{J.~C.} \bibnamefont{Pati}},
  \bibinfo{journal}{Phys. Rev. D} \textbf{\bibinfo{volume}{11}},
  \bibinfo{pages}{566} (\bibinfo{year}{1975}{\natexlab{a}}).

\bibitem[{\citenamefont{Mohapatra and
  Pati}(1975{\natexlab{b}})}]{Mohapatra:1974gc}
\bibinfo{author}{\bibfnamefont{R.~N.} \bibnamefont{Mohapatra}}
  \bibnamefont{and} \bibinfo{author}{\bibfnamefont{J.~C.} \bibnamefont{Pati}},
  \bibinfo{journal}{Phys. Rev. D} \textbf{\bibinfo{volume}{11}},
  \bibinfo{pages}{2558} (\bibinfo{year}{1975}{\natexlab{b}}).

\bibitem[{\citenamefont{Senjanovic and Mohapatra}(1975)}]{Senjanovic:1975rk}
\bibinfo{author}{\bibfnamefont{G.}~\bibnamefont{Senjanovic}} \bibnamefont{and}
  \bibinfo{author}{\bibfnamefont{R.~N.} \bibnamefont{Mohapatra}},
  \bibinfo{journal}{Phys. Rev. D} \textbf{\bibinfo{volume}{12}},
  \bibinfo{pages}{1502} (\bibinfo{year}{1975}).

\bibitem[{\citenamefont{He et~al.}(1991)\citenamefont{He, Joshi, Lew, and
  Volkas}}]{He:1990pn}
\bibinfo{author}{\bibfnamefont{X.~G.} \bibnamefont{He}},
  \bibinfo{author}{\bibfnamefont{G.~C.} \bibnamefont{Joshi}},
  \bibinfo{author}{\bibfnamefont{H.}~\bibnamefont{Lew}}, \bibnamefont{and}
  \bibinfo{author}{\bibfnamefont{R.~R.} \bibnamefont{Volkas}},
  \bibinfo{journal}{Phys. Rev. D} \textbf{\bibinfo{volume}{43}},
  \bibinfo{pages}{22} (\bibinfo{year}{1991}).

\bibitem[{\citenamefont{Moroi and Niki}(2023)}]{Moroi:2022qwz}
\bibinfo{author}{\bibfnamefont{T.}~\bibnamefont{Moroi}} \bibnamefont{and}
  \bibinfo{author}{\bibfnamefont{A.}~\bibnamefont{Niki}},
  \bibinfo{journal}{JHEP} \textbf{\bibinfo{volume}{05}}, \bibinfo{pages}{016}
  (\bibinfo{year}{2023}), \eprint{2205.11766}.

\bibitem[{\citenamefont{Davoudiasl et~al.}(2012)\citenamefont{Davoudiasl, Lee,
  and Marciano}}]{Davoudiasl:2012ag}
\bibinfo{author}{\bibfnamefont{H.}~\bibnamefont{Davoudiasl}},
  \bibinfo{author}{\bibfnamefont{H.-S.} \bibnamefont{Lee}}, \bibnamefont{and}
  \bibinfo{author}{\bibfnamefont{W.~J.} \bibnamefont{Marciano}},
  \bibinfo{journal}{Phys. Rev. D} \textbf{\bibinfo{volume}{85}},
  \bibinfo{pages}{115019} (\bibinfo{year}{2012}), \eprint{1203.2947}.

\end{thebibliography}
\end{raggedright}

\onecolumngrid

\appendix

\section{Differential Cross-Section and Signal} \label{app:signal}

The polarization and spin-averaged squared amplitude is given by
\begin{align}
    \overline{|{\cal M}|^2} = e^4 g_{A'}^2 \frac{Z^2F(t)^2}{t^2}\overline{|{\cal A}|^2},    
\end{align}

The normalized spin-averaged squared amplitude $\overline{|{\cal A}|^2}$ can be written in terms of the Mandelstam variables $\tilde{s}$, $t$, $\tilde{u}$, and $t_2$, defined by
\begin{align}
    \tilde{s} &= (p'+k)^2 - m_e^2\\
    \tilde{u} &= (p-k)^2 - m_e^2\\
    t_2 &= (p' - p)^2\\
    t &= -q^2
\end{align}
which satisfy $\tilde{s} + \tilde{u} + t_2 + t = m_{A'}^2$. With these variables, the squared amplitude can be written as 
\begin{align}
    \overline{|{\cal A}|^2}& = 2\frac{\tilde{s}^2+\tilde{u}^2}{\tilde{s}\tilde{u}}P^2 - \frac{8t}{\tilde{s}\tilde{u}}\left[(P\cdot p)^2 + (P\cdot p')^2 + \frac{t_2 + m_{A'}^2}{2}P^2 \right]\nonumber\\ &+ 2\frac{(\tilde{s}+\tilde{u})^2}{\tilde{s}^2\tilde{u}^2}(m_{A'}^2 + 2m_e^2)\left[P^2 t - 4\left(\frac{\tilde{u}P\cdot p + \tilde{s}P\cdot p'}{\tilde{s}+\tilde{u}}\right)^2\right].
\end{align}

The ion-frame production cross-section is then given by 
\begin{align}
    \frac{d\sigma}{dx\,d(\cos{\theta_k})} &= \frac{1}{512\pi^3 M^2}\frac{|{\bf k}|E}{V|{\bf p}|} \int_{t_-}^{t_+}dt\int_0^{2\pi}\frac{d\phi_q}{2\pi}\overline{|{\cal M}|^2}.
\end{align}
Here, $E$ is the energy of the initial-state electron, $E_k$ is the energy of the vector boson, $V = |{\bf p} - {\bf k}|$, $\cos{\theta_k}$ is the angle the vector boson makes with the beam axis, and $x = E_k/E$ is the fraction of energy transferred to the vector boson. In the lab frame, the expected kinematics are $E_{e^-}^{\rm lab} = 18\,{\rm GeV}$ and $E_{\rm ion}^{\rm lab} = 197\times110\,{\rm GeV} = 21.67\,{\rm TeV}$,\cite{AbdulKhalek:2021gbh} corresponding to a boost of $\gamma_{I} = E_{\rm ion}/M \approx 118$ and a velocity of $v_{I} \approx 0.999964$. Hence, in the rest-frame of the ion, the energy of the initial electron is $E = \gamma_{I}(E_{e^-}^{\rm lab} + v_{I}|{\bf p}^{\rm lab}_{e^-}|) \approx 4250\,{\rm GeV}$.

It is more useful to integrate the cross-section in the lab frame, because limits of integration coincide more directly with detector requirements. In what follows, we will denote quantities in the lab-frame with a ``lab'' superscript. Then, boosting the vector boson's four-momentum from the lab frame to the ion frame yields the equation
\begin{align}
    k^\mu =  \left(E_k, |{\bf k}|\cos{\theta_k}, 0, |{\bf k}|\sin{\theta_k}\right) &= \left(\gamma_{I}\left(E_k^{\rm lab}+v_I|{\bf k}^{\rm lab}|\cos{\theta_k^{\rm lab}}\right), \gamma_I\left(|{\bf k}^{\rm lab}|\cos{\theta_k^{\rm lab}} + v_I E_k^{\rm lab}\right), 0, |{\bf k}^{\rm lab}|\sin{\theta_k^{\rm lab}}\right)
\end{align}
Let $\gamma^{\rm lab}_k \equiv E_k^{\rm lab}/m_{A'}$ and $v^{\rm lab}_k \equiv |{\bf k}^{\rm lab}|/E_k^{\rm lab}$. Additionally, if the pseudorapidity is $\eta_k^{\rm lab} = -\log(\tan(\theta_k^{\rm lab} / 2))$, then $\cos{\theta_k^{\rm lab}} = \tanh{\eta_k^{\rm lab}}$ and $\sin{\theta_k^{\rm lab}} = {\rm sech\,}{\eta_k^{\rm lab}}$. With these substitutions, the equation for the zeroth component reads
\begin{align}
    xE &= \gamma_I \gamma_k^{\rm lab}m_{A'}\left(1 + v_I v_k^{\rm lab}\tanh{\eta_k^{\rm lab}}\right),
\end{align}
and the quotient of the third and first components yields
\begin{align}
    \tan{\theta_k} &= \frac{v_k^{\rm lab}{\rm sech\,}\eta_k^{\rm lab}}{\gamma_I( v_k^{\rm lab}\tanh{\eta_k^{\rm lab}}+ v_I)}.
\end{align}
The corresponding Jacobian determinant of the transformation from $(x, \cos{\theta_k})$ to $(\gamma_k^{\rm lab}, \eta_k^{\rm lab})$ is
\begin{align}
    \left|\frac{\partial x}{\partial \gamma_k^{\rm lab}}\frac{\partial (\cos{\theta_k})}{\partial \eta_k^{\rm lab}} - \frac{\partial (\cos{\theta_k})}{\partial \gamma_k^{\rm lab}}\frac{\partial x}{\partial \eta_k^{\rm lab}}\right| &= \frac{m_{A'}}{E}{\rm sech\,}\eta_k^{\rm lab}\sin{\theta_k}.
\end{align}
Hence, the differential cross-section with respect to boost and pseudorapidity in the lab frame is
\begin{align}
    \frac{d\sigma}{d\gamma_k^{\rm lab}d\eta_k^{\rm lab}} &= \frac{m_{A'}}{E}{\rm sech\,}\eta_k^{\rm lab}\sin{\theta_k}\frac{d\sigma}{dx\,d(\cos{\theta_k})}.
\end{align}

The decay width of a given hidden vector $A'$ can be written as
\beq
    \Gamma_{A'} = \sum_\nu Q_\nu^2\Gamma_{\bar{\nu}\nu} + \sum_\ell Q_\ell^2\Gamma_{\bar{\ell}\ell} + \Gamma_{\rm had},
\eeq
where for one flavor of left-handed neutrino,
\begin{align}
    \Gamma_{\bar{\nu}\nu} &= \frac{g_{A'}^2}{24\pi}m_{A'}\\
\intertext{for charged fermions,}
    \Gamma_{\bar{f}f} &= \frac{g_{A'}^2}{12\pi}m_{A'}\left(1 + \frac{2m_f^2}{m_{A'}^2}\right)\sqrt{1 - \frac{4m_{f}^2}{m_{A'}^2}},\\
\intertext{and for hadrons}
    \Gamma_{\rm had} &= \frac{\sum_{q}Q_q^2\Gamma_{\bar{q}q}}{\frac{4}{9}\sum_u{\Gamma_{\bar{u}u}} + \frac{1}{9}\sum_d{\Gamma_{\bar{d}d}}} \Gamma_{\bar{\mu}\mu}R(m_{A'}^2).
\end{align}
Here, $R(m_{A'}^2) = \sigma(e^+e^- \rightarrow {\rm hadrons})/\sigma(e^+e^- \rightarrow \mu^+\mu^-)$ is the experimentally determined $R$-ratio. The detailed charge assignments $Q_i$ which we use for the dark photon, $B-L$ and leptophilic gauge boson follow \cite{Bauer:2018onh}.

\section{Distance of Closest Approach}\label{app:dca}

The distance-of-closest-approach (DCA) for a particle is defined as the minimal distance between the particle's reconstructed trajectory and the primary vertex. For particles produced in a displaced decay (such as the lepton pairs in our process), the trajectory is reconstructed assuming the particle was produced at a much earlier time. The transverse DCA (${\rm DCA}_{\rm 2D}$) is then given by projecting the DCA onto the transverse plane.

\begin{figure}
    \centering
    \includegraphics[width=0.5\linewidth]{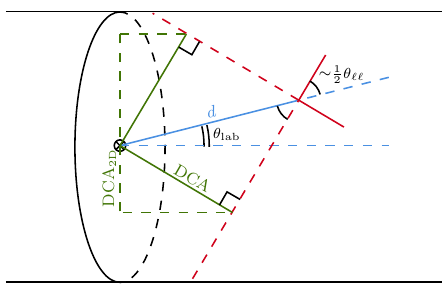}
    \caption{Schematic showing the definition of the transverse DCA. The dark boson travels a distance $d$ in the detector (shown in blue), before decaying into a pair of leptons (shown in red). The transverse DCA of these leptons is found by tracing their trajectories backwards (red dashed lines), then projecting onto the transverse plane. This diagram shows the scenario in which the leptons decay with a polar angle of $\varphi_- = 0$ and $\varphi_+ = \pi$, but the polar angle can in general be anywhere between $0$ and $\pi$.}
    \label{fig:dca}
\end{figure}

A diagram showcasing our estimation of the DCA in terms of the kinematic properties of the dark boson can be seen in Fig.~\ref{fig:dca}. For simplicity, we take the angle of the lepton with respect to the vector to be half the opening angle between the two leptons, $\theta_{\ell\ell}/2$; a more detailed study of the event kinematics shows that this situation is typical. To estimate the DCA, we assume that the leptons follow a straight-line path backward from where they were produced. To justify this assumption, note that in the EIC, the magnetic field is expected to be $B \sim 1~{\rm T}$, and the energy $\gamma m_e$ of the electron decay products is typically much larger than the mass $m_{A'}$ of the dark photon. Hence, the radius of the final-state electron or positron's trajectory is given by
\beq
    R = \frac{\gamma m_e v}{eB} \gg \left(\frac{m_{A'}}{\rm 100\,MeV}\right){\rm m}.
\eeq
Hence, for the parameter region of our study, it is on the order of a few meters, which is much larger than the minimum DCA requirement.

Note that the schematic is a special case, in which the leptons are traveling with a polar angle $\varphi = 0$ or $\varphi = \pi$ with respect to the detector, with the DCA drawn for the lepton with $\varphi = 0$. Regardless of this angle, for a straight-line trajectory, the DCA can be related to the travel length by
\begin{equation}
    {\rm DCA} = d \sin{(\theta_{\ell\ell}/2)}.
\end{equation}
The relationship between the DCA and the transverse DCA depends on the polar angle $\varphi$. In general, it is given by
\begin{equation}
    {\rm DCA}_{\rm 2D} = {\rm DCA}\sqrt{1 - \left(\cos{\theta_{\rm lab}}\sin{(\theta_{\ell\ell/2})} + \sin{\theta_{\rm lab}}\cos{(\theta_{\ell\ell}/2)}\cos{\varphi}\right)^2}.
\end{equation}
For the cases $\varphi = 0$ and $\varphi = \pi$, the transverse DCA is given by
\begin{equation}
    {\rm DCA}_{\rm 2D} = {\rm DCA}\cos{(\theta_{\rm lab} \pm \theta_{\ell\ell}/2)}.
\end{equation}
Depending on the angle $\varphi$, the transverse DCA can be anywhere between these values. For simplicity, we assume that the average transverse DCA is given by the average of these maximum and minimum values, which yields
\begin{align}
    \overline{{\rm DCA}_{\rm 2D}} &= {\rm DCA}\cos{\theta_{\rm lab}}\cos{(\theta_{\ell\ell}/2)}\\
        &= \frac{1}{2}d \cos{\theta_{\rm lab}}\sin{\theta_{\ell\ell}}.
\end{align}
The angle $\theta_{\ell\ell}$ can be related to the boost $\gamma$ through four-momentum conservation:
\begin{align}
    \gamma m_{A'} &= 2\gamma_{\ell}m_{\ell},\\
    \gamma v m_{A'} &= 2\gamma_{\ell} v_{\ell} m_{\ell}\cos{(\theta_{\ell\ell}/2)}.
\end{align}
Solving for $\sin{\theta_{\ell\ell}}$ yields
\begin{align}
    \sin{\theta_{\ell\ell}} &= \frac{2\gamma v m_{A'}}{\gamma^2 m_{A'}^2 - 4m_{\ell}^2}\sqrt{m_{A'}^2 - 4m_{\ell}^2} \approx \frac{2v}{\gamma},
\end{align}
since $m_{\ell} = m_e \ll m_{A'}$ for this study. The minimum resolvable value of transverse DCA for particles at the EIC is cited to be around $100\,{\rm \mu m}$ \cite{Adkins:2022jfp}. With this, the minimum distance the dark photon can travel before being considered displaced is
\begin{align}
    d_{\rm min} \approx \frac{\gamma}{v\cos{\theta_{\rm lab}}}(100\,{\rm \mu m}).
\end{align}

\clearpage

\end{document}